\documentclass[preprint]{JHEP}

\usepackage{epsfig}
\usepackage{tabularx}
\usepackage{graphicx}
\usepackage{amssymb,amsfonts,amsmath}

\title{Dark matter in supersymmetric models with axino LSP in Randall-Sundrum II brane model}

\author{Jin U Kang$^1$,$^2$ and Grigoris Panotopoulos$^2$ \\

$^1$Department of Physics, Kim Il Sung University, \\
Pyongyang, Democratic People's Republic of Korea \\

$^2$ASC, Department of Physics LMU,\\
Theresienstr. 37, 80333 Munich, Germany\\

{\tt E-mail: Jin.U.Kang@physik.uni-muenchen.de, Grigoris.Panotopoulos@physik.uni-muenchen.de}}

\abstract{The axino dark matter hypothesis in RSII brane model is studied. Within the framework of CMSSM we assume that the lightest neutralino or stau is the NLSP, and that the axino production has a single contribution from the NLSP decay. It is found that the axino can play the role of dark matter in the universe and we determine what the axino mass should be for different values of the five-dimensional Planck mass. An upper bound is obtained for the latter.}

\begin{document}


\setcounter{equation}{0}

\section{Introduction}

There are good theoretical reasons for which particle physics proposes that new exotic particles must exist. In particular, the strong CP problem and the hierarchy problem motivate symmetries and particles beyond the standard model of particle physics. On one hand, supersymmetry (SUSY) is an ingredient that appears in many theories for physics beyond the standard model. SUSY solves the hierarchy problem and predicts that every particle we know should be escorted by its superpartner. In order for the supersymmetric solution of the hierarchy problem to work, it is necessary that the SUSY becomes manifest at relatively low energies, less than a few $TeV$, and therefore the required superpartners must have masses below this scale (for supersymmetry and supergravity see e.g.~\cite{nilles}). On the other hand, the strong CP problem can be solved naturally by implementing the Peccei-Quinn (PQ) mechanism~\cite{quinn}. An additional global $U(1)$ symmetry referred to as PQ symmetry broken spontaneously at the PQ scale can explain the smallness of the CP-violating $\Theta$-vacuum term in quantum chromodynamics (QCD). The pseudo Nambu-Goldstone boson associated with this spontaneous symmetry breaking is the axion~\cite{wilczek}, which has not yet been detected. Axinos, the superpartners of axions, are special because they have unique properties: They are very weekly interacting and their mass can span a wide range, from very small ($ \sim eV$)to large ($ \sim GeV$) values. What is worth stressing is that, in contrast to the neutralino and the gravitino, axino mass does not have to be of the order of the SUSY breaking scale in the visible sector, $M_{SUSY} \sim 100GeV-1TeV$.

One of the theoretical problems in modern cosmology is to understand the nature of cold dark matter in the universe. There are good reasons, both observational and theoretical, to suspect that a fraction of $0.22$ of the energy density in the universe is in some unknown ``dark'' form. Many lines of reasoning suggest that the dark matter consists of some new, as yet undiscovered, massive particle which experiences neither electromagnetic nor color interactions. In SUSY models which are realized with R-parity conservation the lightest supersymmetric particle (LSP) is stable. A popular cold dark matter candidate is the LSP, provided that it is electrically and color neutral. Certainly the most theoretically developed LSP is the lightest neutralino~\cite{neutralino}. However, there are other dark matter candidates as well, for example the gravitino~\cite{gravitino, steffen1} and the axino~\cite{axino, steffen2}, the superpartner of axion~\cite{wilczek} which solves the QCD problem via the Peccei-Quinn mechanism~\cite{quinn}. In this article we work in the framework of Randall-Sundrum type II brane model (RSII), we assume that the axino is the LSP and address the question whether the axino can play the role of dark matter in the universe, and for which range for axino mass and five-dimensional Planck mass.

Our work is organized as follows: The article consists of four section, of which this introduction is the first. In the second section we present the theoretical framework, while in section 3 we show the results of our analysis. Finally we conclude in the last section.

\section{The theoretical framework}

\subsection{The brane model}

Over the last years the brane-world models have been attracting a lot of
attention as a novel higher-dimensional theory. Brane models are inspired from
M/string theory and although they are not yet derivable from the fundamental
theory, at least they contain the basic ingredients, like extra dimensions,
higher-dimensional objects (branes), higher-curvature corrections to gravity
(Gauss-Bonnet) etc. Since string theory claims to give us a fundamental
description of nature it is important to study what kind of cosmology it
predicts. Furthermore, despite the fact that supersymmetric dark matter has been
analyzed in standard four-dimensional cosmology, it is challenging to discuss
it in alternative gravitational theories as well. Neutralino dark matter in brane cosmology has been studied in~\cite{okada1}, while axino dark matter in brane-world cosmology has been studied in~\cite{panotop}.

In brane-world models it
is assumed that the standard model particles are confined on a 3-brane while
gravity resides in the whole higher dimensional spacetime. The model first
proposed by Randall and Sundrum (RSII)~\cite{rs}, is a simple and interesting
one, and its cosmological evolutions have been intensively
investigated. An incomplete list can be seen e.g. in~\cite{langlois}. In the present
work we would like to study
axino dark matter in the framework of RSII model. According to that
model, our 4-dimensional
universe is realized on the 3-brane with a positive tension located at the UV
boundary of 5-dimensional AdS spacetime. In the bulk there is just a
cosmological constant $\Lambda_{5}$, whereas on the brane there is matter with
energy-momentum tensor $\tau_{\mu \nu}$. Also, the five dimensional Planck
mass is denoted by $M_{5}$ and the brane tension is denoted by $T$.

If Einstein's equations hold in the five dimensional bulk, then it has been shown in \cite{shiromizu} that the effective four-dimensional Einstein's equations induced on the brane can be written as
\begin{equation}
G_{\mu \nu}+\Lambda_{4} g_{\mu \nu}=\frac{8 \pi}{m_{pl}^2} \tau_{\mu \nu}+(\frac{1}{M_{5}^3})^2 \pi_{\mu \nu}-E_{\mu \nu}
\end{equation}
where $g_{\mu \nu}$ is the induced metric on the brane, $\pi_{\mu \nu}=\frac{1}{12} \: \tau \: \tau_{\mu \nu}+\frac{1}{8} \: g_{\mu \nu} \: \tau_{\alpha \beta} \: \tau^{\alpha \beta}-\frac{1}{4} \: \tau_{\mu \alpha} \: \tau_{\nu}^{\alpha}-\frac{1}{24} \: \tau^2 \: g_{\mu \nu}$, $\Lambda_{4}$ is the effective four-dimensional cosmological constant, $m_{pl}$ is the usual four-dimensional Planck mass and $E_{\mu \nu} \equiv C_{\beta \rho \sigma} ^\alpha \: n_{\alpha} \: n^{\rho} \: g_{\mu} ^{\beta} \: g_{\nu} ^{\sigma}$ is a projection of the five-dimensional Weyl tensor $C_{\alpha \beta \rho \sigma}$, where $n^{\alpha}$ is the unit vector normal to the brane.
The tensors $\pi_{\mu \nu}$ and $E_{\mu \nu}$ describe the influence of the bulk in brane dynamics. The five-dimensional quantities are related to the corresponding four-dimensional ones through the relations
\begin{equation}
m_{pl}=4 \: \sqrt{\frac{3 \pi}{T}} \: M_{5}^3
\end{equation}
and
\begin{equation}
\Lambda_{4}=\frac{1}{2 M_{5}^3} \left( \Lambda_{5}+\frac{T^2}{6 M_{5}^3} \right )
\end{equation}
In a cosmological model in which the induced metric on the brane $g_{\mu \nu}$ has the form of  a spatially flat Friedmann-Robertson-Walker model, with scale factor $a(t)$, the Friedmann-like equation on the brane has the generalized form~\cite{langlois}
\begin{equation}
H^2=\frac{\Lambda_{4}}{3}+\frac{8 \pi}{3 m_{pl}^2}  \rho+\frac{1}{36 M_{5}^6} \rho^2+\frac{C}{a^4}
\end{equation}
where $C$ is an integration constant arising from $E_{\mu \nu}$. The cosmological constant term and the term linear in $\rho$ are familiar from the four-dimensional conventional cosmology. The extra terms, i.e the ``dark radiation'' term and the term quadratic in $\rho$, are there because of the presence of the extra dimension. Adopting the Randall-Sundrum fine-tuning
\begin{equation}
\Lambda_{5}=-\frac{T^2}{6 M_{5}^3}
\end{equation}
the four-dimensional cosmological constant vanishes. In addition, the dark radiation term is severely constrained by the success of the Big Bang Nucleosynthesis (BBN), since the term behaves like an additional radiation at the BBN era \cite{orito}. So, for simplicity, we neglect the term in the following analysis. The five-dimensional Planck mass is also constrained by the BBN, which is roughly estimated as $M_{5} \geq 10 \: TeV$ \cite{cline}. The generalized Friedmann equation takes the final form
\begin{equation}
H^2=\frac{8 \pi G}{3} \rho \left (1+\frac{\rho}{\rho_0} \right )
\end{equation}
where
\begin{equation}
\rho_0=96 \pi G M_{5}^6
\end{equation}
with $G$ the Newton's constant. One can see that the evolution of the early universe can be divided into two eras. In the low-energy regime $\rho \ll \rho_0$ the first term dominates and we recover the usual Friedmann equation of the conventional four-dimensional cosmology. In the high-energy regime $\rho_0 \ll \rho$ the second term dominates and we get an unconventional expansion law for the universe. In between there is a transition temperature $T_t$ for which $\rho(T_t)=\rho_0$. Once $M_{5}$ is given, the transition temperature $T_{t}$ is determined as
\begin{equation}
T_{t}=1.6 \times 10^{7} \: \left ( \frac{100}{g_{eff}} \right )^{1/4} \: \left ( \frac{M_{5}}{10^{11} \: GeV} \right )^{3/2} \: GeV
\end{equation}
where $g_{eff}$ counts the total number of relativistic degrees of freedom.

\subsection{The particle physics model}

The extension of standard model (SM) of particle physics based on SUSY is the minimal supersymmetric standard model (MSSM)~\cite{mssm}. It is a supersymmetric gauge theory based on the SM gauge group with the usual representations (singlets, doublets, triplets) and on $\mathcal{N}=1$ SUSY. Excluding gravity, the massless representations of the SUSY algebra are a chiral and a vector supermultiplet. The gauge bosons and the gauginos are members of the vector supermultiplet, while the matter fields (quarks, leptons, Higgs) and their superpartners are members of the chiral supermultiplet. The Higgs sector in the MSSM is enhanced compared to the SM case. There are now two Higgs doublets, $H_u, H_d$, for anomaly cancelation requirement and for giving masses to both up and down quarks. After electroweak symmetry breaking we are left with five physical Higgs bosons, two charged $H^{\pm}$ and three neutral $A,H,h$ ($h$ being the lightest). Since we have not seen any superpartners yet SUSY has to be broken. In MSSM, SUSY is softly broken by adding to the Lagrangian terms of the form
\begin{itemize}
\item Mass terms for the gauginos $\tilde{g}_i$, $M_1, M_2, M_3$
\begin{equation}
M \tilde{g} \tilde{g}
\end{equation}
\item Mass terms for sfermions $\tilde{f}$
\begin{equation}
m_{\tilde{f}}^2 \tilde{f}^{\dag} \tilde{f}
\end{equation}
\item Masses and bilinear terms for the Higgs bosons $H_u, H_d$
\begin{equation}
m_{H_u}^2 H_u^{\dag} H_u+m_{H_d}^2 H_d^{\dag} H_d+B \mu (H_u H_d + h.c.)
\end{equation}
\item Trilinear couplings between sfermions and Higgs bosons
\begin{equation}
A Y \tilde{f}_1 H \tilde{f}_2
\end{equation}
\end{itemize}
In the unconstrained MSSM there is a huge number of unknown parameters~\cite{parameters} and thus little predictive power. However, the Constrained MSSM (CMSSM) or mSUGRA~\cite{msugra} is a framework with a small controllable number of parameters, and thus with much more predictive power. In the CMSSM there are four parameters, $m_0, m_{1/2}, A_0, tan \beta$, which are explained below, plus the sign of the $\mu$ parameter from the Higgs sector. The magnitude of $\mu$ is determined by the requirement for a proper electroweak symmetry breaking, its sign however remains undetermined. We now give the explanation for the other four parameters of the CMSSM
\begin{itemize}
\item Universal gaugino masses
\begin{equation}
M_1(M_{GUT})=M_2(M_{GUT})=M_3(M_{GUT})=m_{1/2}
\end{equation}
\item Universal scalar masses
\begin{equation}
m_{\tilde{f}_i}(M_{GUT})=m_0
\end{equation}
\item Universal trilinear couplings
\begin{equation}
A_{i j}^u(M_{GUT}) = A_{i j}^d(M_{GUT}) = A_{i j}^l(M_{GUT}) = A_0 \delta_{i j}
\end{equation}
\item
\begin{equation}
tan \beta \equiv \frac{v_1}{v_2}
\end{equation}
where $v_1, v_2$ are the vevs of the Higgs doublets and $M_{GUT} \sim 10^{16}~GeV$ is the Grand Unification scale.
\end{itemize}

\section{Analysis and results}

We consider eight benchmark models (shown in Table~1 and Table~2) for natural values of $m_0, m_{1/2}$, representative values of $tan \beta$ and fixed $A_0=0, \mu >0$.
In these models the lightest neutralino (denoted by $\chi$) or the lightest stau (denoted by $\tilde{\tau}$) is the lightest of the usual superpartners and thus the NLSP.
Furthermore the following experimental constraints (for the lightest Higgs mass and a rare decay)~\cite{precision, Yao:2006px} are satisfied

\begin{eqnarray}
m_h & > & 114.4~GeV \\
BR(b \rightarrow s \gamma) & = & (3.39_{-0.27}^{+0.30}) \times 10^{-4}
\end{eqnarray}

\begin{table}
\begin{center}
\begin{tabular}{|c|c|c|c|c|c|}
\hline
 Model & $m_0 \: (GeV)$ & $m_{1/2} \: (GeV)$ & $tan \beta$ & $m_{\chi} \: (GeV)$ & $\Omega_{\chi} h^2$  \\ \hline
 A & 200 & 500 & 15 & 205.42 & 0.64 \\ \hline
 B & 400 & 800 & 25 & 337.95 & 1.82 \\\hline
 C & 1000 & 600 & 30 & 252.41 & 7.37 \\ \hline
 D & 350 & 450 & 20 & 184.46 & 1.2 \\ \hline
\multicolumn{6}{l}{Table 1: Four benchmark models considered in the analysis for the neutralino NLSP case.}
\end{tabular}
\end{center}
\end{table}

\begin{table}
\begin{center}
\begin{tabular}{|c|c|c|c|c|c|}
\hline
 Model & $m_0 \: (GeV)$ & $m_{1/2} \: (GeV)$ & $tan \beta$ & $m_{\tilde{\tau}} \: (GeV)$ & $\Omega_{\tilde{\tau}} h^2$  \\ \hline
 E & 50 & 500 & 10 & 187.78 & 0.0088 \\ \hline
 F & 60 & 600 & 11 & 223.47 & 0.012 \\\hline
 G & 70 & 700 & 12 & 258.95 & 0.017 \\ \hline
 H & 100 & 800 & 15 & 295.93 & 0.022 \\ \hline
\multicolumn{6}{l}{Table 2: Four benchmark models considered in the analysis for the stau NLSP case.}
\end{tabular}
\end{center}
\end{table}

At this point we remark that any viable model should also satisfy two more mass limits~\cite{Yao:2006px}
\begin{eqnarray}
m_{\tilde{\tau}_1} & > & 81.9~GeV \\
m_{\tilde{\chi}_1^{\pm}} & > & 94~GeV
\end{eqnarray}
However, in the models we consider here the NLSP mass is at least $m_{NLSP} \simeq 184~GeV$ and therefore further imposing limits of ${\cal O}(100~GeV)$ on other sparticles is meaningless.

The SUSY spectrum (as well as the Higgs bosons masses) and the neutralino relic density have been computed using the web site~\cite{kraml}, and the top quark mass is fixed to $m_t=172.7~GeV$~\cite{cdf}. Furthermore, following~\cite{okada2} for the stau relic density we have made use of the simple formula
\begin{equation}
\Omega_{\tilde{\tau}}h^2=\left( \frac{m_{\tilde{\tau}}}{2~TeV} \right )^2
\end{equation}

Before proceeding any further a couple of remarks are in order. First, we mention that in principle the saxion (a scalar field in the same supermultiplet with axion and axino) could have important cosmological consequences. Here, however, we shall assume that the saxion mass is such that its cosmological consequences are negligible. This kind of assumption was also made in~\cite{steffen2}. Furthermore, in two previous works~\cite{steffen2, panotop} the axino dark matter in standard and brane cosmology was considered, in which the axino thermal production only was taken into account. There it was found that the reheating temperature (in standard cosmology) or the transition temperature (in brane cosmology) had to be bounded from above, $T_{R,t} \leq 10^6~GeV$. However, at this temperature the strong coupling constant is of the order one, $g_s \sim 1$, a fact which may render the whole discussion invalid\footnote{We would like to thank F.~D.~Steffen for pointing this out.}. That is why in the present work we have chosen to only consider the non-thermal production from the NLSP decay. If we restrict ourselves to small $M_5$ or $T_t$ we can neglect the thermal production mechanism as being negligible compared to the non-thermal production mechanism. Finally, in principle one should also impose the BBN constraints (see e.g.~\cite{kohri}) if the NLSP decays after BBN time. However in the axino dark matter case the BBN constraints are easily avoided because the NSLP has a relatively short lifetime and decays well before BBN~\cite{axino}a.

\begin{figure}
\centerline{\epsfig{figure=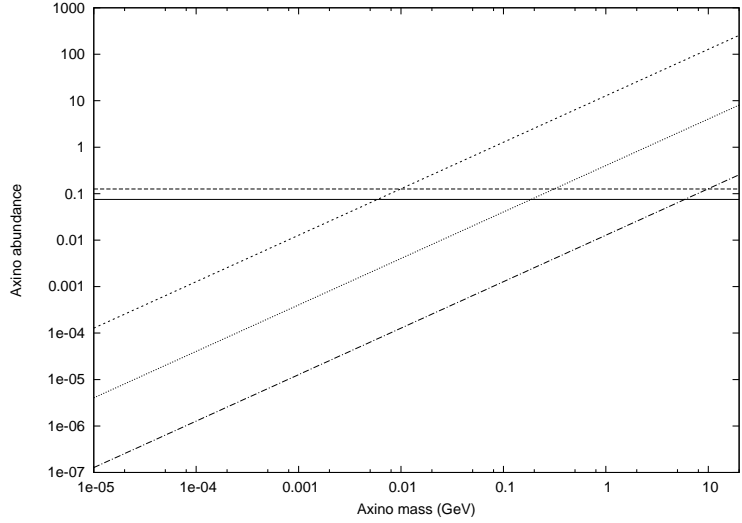,height=7cm,angle=0}}
\caption{Axino abundance versus axino mass for several values of the five-dimensional Planck mass (Neutralino NLSP, benchmark model A ). The strip around $0.1$ is the allowed range for cold dark matter. Values of $M_5$ used are $10^4~GeV$, $10^5~GeV$, and $10^6~GeV$ from top to bottom.}
\end{figure}

\begin{figure}
\centerline{\epsfig{figure=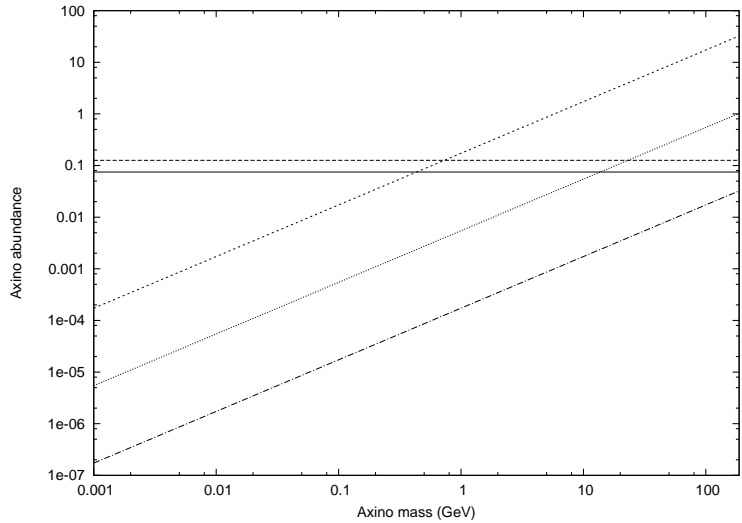,height=7cm,angle=0}}
\caption{Same as figure $1$, but for the stau NLSP case, benchmark model E.}
\end{figure}

\begin{figure}
\centerline{\epsfig{figure=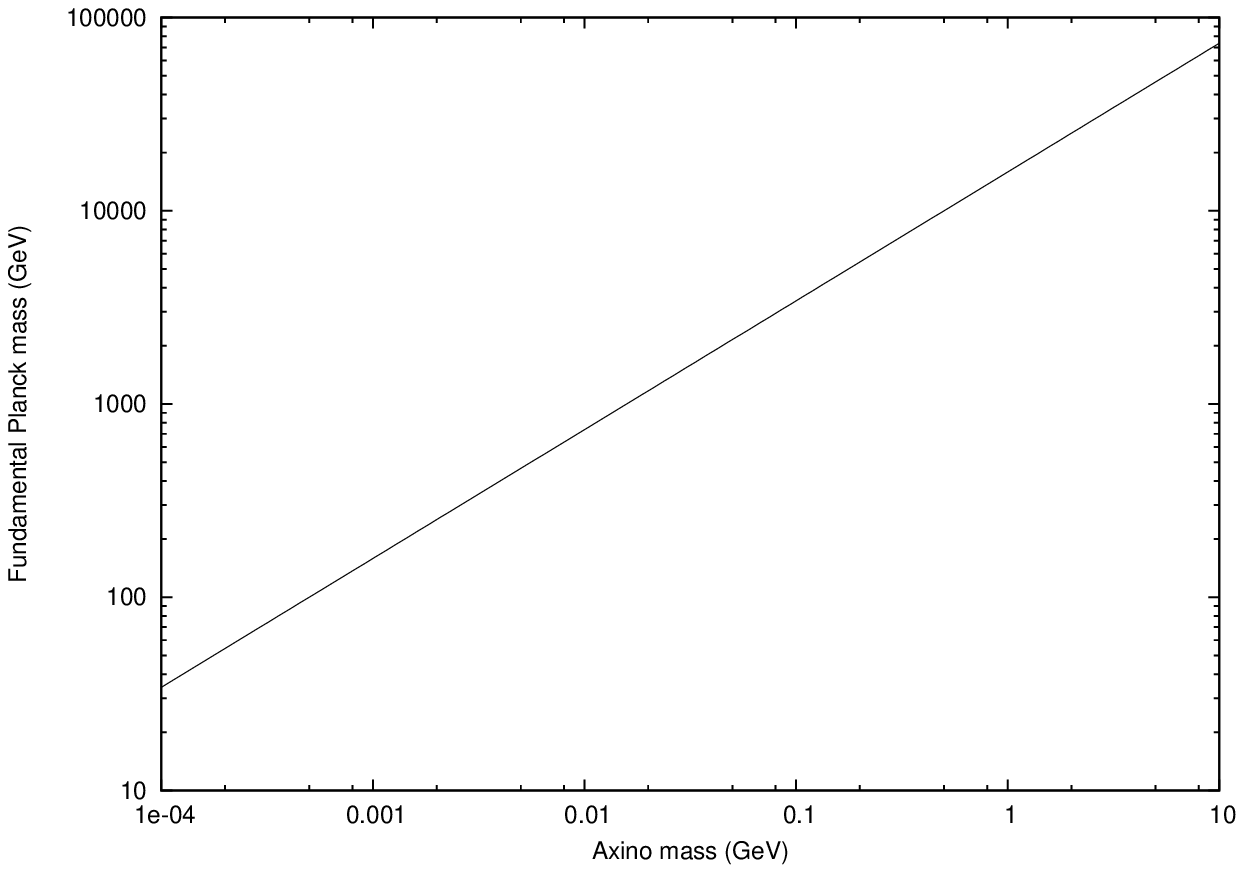,height=7cm,angle=0}}
\caption{$M_5$ versus $m_{\tilde{\alpha}}$ for the stau NLSP case. For the observational value $\Omega_{\tilde{\alpha}}^{(b)} h^2=0.1$ one can see the upper bound on $M_5$ imposing that $m_{\tilde{\alpha}} \leq 10~GeV$.}
\end{figure}

For the axino abundance we take into account the non-thermal production (NTP)
and we impose the WMAP constraint for cold dark matter~\cite{wmap}
\begin{equation}
0.075 < \Omega_{cdm} h^2=\Omega_{\tilde{\alpha}} h^2 < 0.126
\end{equation}
In the NTP case the contribution to the axino abundance comes from the decay of the NLSP
\begin{equation}
\Omega_{\tilde{\alpha}} h^2 = \frac{m_{\tilde{\alpha}}}{m_{NLSP}} \: \Omega_{NLSP} h^2
\end{equation}
with $m_{\tilde{\alpha}}$ the axino mass, $m_{NLSP}$ the mass of the NLSP and $\Omega_{NLSP} h^2$ the NLSP abundance had it did not decay into the axino.

Now we need to take into account the effect of the novel law for expansion of the universe. The relic density of a particle of mass $m$ is modified as follows~\cite{okada3}
\begin{equation}
\frac{\Omega^{(b)}}{\Omega^{(s)}}=0.54 \: \frac{x_t}{x_{d}^{(s)}}
\end{equation}
in the limit $x_t \gg x_d$ and in the S-wave approximation, where the index b stands for "brane", the index s stands for "standard", $x_t=m/T_t$ and $x_d=m/T_d$, with $T_t$ the transition temperature and $T_d$ the decoupling temperature of the particle of mass $m$. In standard cosmology $x_d^{(s)} \simeq 30$. In a given particle physics model the axino abundance in terms of $M_5$ and $m_{\tilde{\alpha}}$ is given by
\begin{equation}
\Omega_{\tilde{\alpha}} h^2 = 0.9 \times 10^{7.5} \: \left ( \frac{m_{\tilde{\alpha}}}{GeV} \right ) \: \Omega_{NLSP}^{(s)} h^2 \: \left ( \frac{M_5}{GeV} \right )^{-3/2}
\end{equation}

For each benchmark model we have obtained plots (for example we show the plots for models A and E, for the rest of the models there are similar plots) which show the axino abundance $\Omega_{\tilde{\alpha}} h^2$ as a function of the axino mass $m_{\tilde{\alpha}}$ for several values of the five-dimensional Planck mass $M_5$.  Figure $1$ corresponds to the neutralino NLSP case (the values that we have used are $10^4~GeV$, $10^5~GeV$, and $10^6~GeV$ from top to bottom), while figure $2$ corresponds to the stau NLSP case (same values of $M_5$). We see that there is always one allowed range for the axino mass from the milli-$GeV$ range to a few $GeV$. If however $M_5$ is high enough, in the stau NLSP case the axino has to be very heavy. In this case, using the formula above for the axino abundance and imposing the condition that $m_{\tilde{\alpha}} \leq 10~GeV$, it is easy to show that for $\Omega_{NLSP}^{(s)} h^2=0.01$ the fundamental Planck mass is bounded from above, $M_5 \leq 7.4 \times 10^4~GeV$. This can be shown in figure 3.

\section{Conclusions}

We have studied axino dark matter in the brane-world cosmology. The theoretical framework for our work is the CMSSM for particle physics and RS II for gravity, which predicts a generalized Friedmann-like equation for the evolution of the universe. We assume that axino is the LSP and the lightest neutralino or the lightest stau is the NLSP. For the axino abundance we have taken into account the non-thermal production and have imposed the cold dark matter constraint $0.075 < \Omega_{cdm} h^2 < 0.126$. The formula valid in standard four-dimensional cosmology is corrected taking into account the novel expansion law for the universe. We have considered eight benchmark models (four for the neutralino NLSP and four for the stau NLSP case) for natural values of $m_0$ and $m_{1/2}$ and representative values of $tan \beta$. In these models the neutralino or the stau is the lightest of the usual superpartners (and thus the NLSP, since we assume that the axino is the LSP) and experimental constraints are satisfied. For each benchmark model we have produced plots of the axino abundance as a function of the axino mass for several different values of the five-dimensional Planck mass. The obtained plots show that in general the axino can be the cold dark matter in the universe for axino masses from  $0.001~GeV$ up to a few $GeV$. Furthermore, in the stau NLSP case an upper bound on the five-dimensional Planck mass is obtained, $M_5 \leq 7.4 \times 10^{4}~GeV$.

\section*{Acknowledgments}

J.U~K is supported by the German Academic Exchange Service (DAAD), and G.~P. is supported by project "Particle Cosmology".

\end{document}